\documentclass[twocolumn,showpacs,superscriptaddress,nofootinbib,amsmath,asmsym]{revtex4-1}
\usepackage{bbm,color}
\usepackage{graphicx}

\usepackage[ocgcolorlinks,colorlinks=true,linkcolor=blue,citecolor=red]{hyperref}

\newcommand{\bra}[1]{\langle#1|}
\newcommand{\ket}[1]{|#1\rangle}

\newcommand{\ketbra}[2]{|#2\rangle\langle#1|}

\newcommand{\Tr}{\mathrm{Tr}}
\newcommand{\tr}{\mathrm{Tr}}

\newcommand{\identity}{\mathbbm{1}}
\newcommand{\guess}{\mathrm{guess}}
\newcommand{\rA}{\mathrm{A}}
\newcommand{\rB}{\mathrm{B}}
\newcommand{\rE}{\mathrm{E}}
\newcommand{\obs}{\mathrm{obs}}
\newcommand{\proj}[1]{|#1\rangle\langle#1|}

\newcommand{\ie}{{\it{i.e.}~}}

\begin{document}

\title{Optimal randomness certification in the quantum steering and prepare-and-measure scenarios}
\author{Elsa Passaro}
\affiliation{ICFO-Institut de Ciencies Fotoniques, Mediterranean
Technology Park, 08860 Castelldefels (Barcelona), Spain}

\author{Daniel Cavalcanti}
\affiliation{ICFO-Institut de Ciencies Fotoniques, Mediterranean
Technology Park, 08860 Castelldefels (Barcelona), Spain}

\author{Paul Skrzypczyk}
\affiliation{H. H. Wills Physics Laboratory, University of Bristol, Tyndall Avenue, Bristol, BS8 1TL, United Kingdom.}
\affiliation{ICFO-Institut de Ciencies Fotoniques, Mediterranean
Technology Park, 08860 Castelldefels (Barcelona), Spain}

\author{Antonio~Ac\'in}
\affiliation{ICFO-Institut de Ciencies Fotoniques, Mediterranean
Technology Park, 08860 Castelldefels (Barcelona), Spain}
\affiliation{ICREA-Instituci\'o Catalana de Recerca i Estudis Avan\c cats, Lluis Companys 23, 08010 Barcelona, Spain\\
}

\begin{abstract}
Quantum mechanics predicts the existence of intrinsically random processes. Contrary to classical randomness, this lack of predictability can not be attributed to ignorance or lack of control. Here we find the optimal method to quantify the amount of local or global randomness that can be extracted in two scenarios: (i) the quantum steering scenario, where two parties measure a bipartite system in an unknown state but one of them does not trust his measurement apparatus, and (ii) the prepare-and-measure scenario, where additionally the quantum state is known. We use our methods to compute the maximal amount of local and global randomness that can be certified by measuring systems subject to noise and losses and show that local randomness can be certified from a single measurement if and only if the detectors used in the test have detection efficiency higher than $50\%$.
\end{abstract}

\maketitle

One of the most distinct features of quantum mechanics is its intrinsically random character. While in classical mechanics lack of predictability can always be associated to ignorance or lack of control of the probed systems, the rules of quantum physics say that one can not predict the outcome of a measurement even if all the variables of a system are known.  This inherent unpredictability has been exploited in different applications such as quantum random number generation \cite{QRNG} and quantum key distribution \cite{QKD_RMP}.

Recent results have shown that the randomness observed in quantum mechanics can be certified even without relying on
any modelling of the quantum devices used for the generation of the random data. In fact, by analysing the data obtained in experiments involving local measurements on bipartite entangled systems one can prove that no one could have predicted this data in advance whenever a Bell inequality violation is observed~\cite{Bell64,Bell review}. This is called device-independent randomness certification \cite{Colbeck,PAMBMMOHLMM10}. The device-independent approach has the practical advantage that it does not rely on the exact description of the experimental set-up. This is crucial when implementing cryptographic protocols as an adversary can use a mismatch between the theoretical description and the actual implementation of the set-up to fake its performance \cite{hack1,hack2,hack3}. However, device-independent protocols require low levels of noise \cite{Bell review}, which make them very demanding experimentally.

An intermediate scenario is that of quantum steering \cite{Schr,WJD07}. It refers to the case where two parties, say Alice and Bob, apply local measurements on an unknown bipartite system. While one of them, Bob, has complete knowledge of his measurement apparatuses, Alice does not, and treats her measuring device as a black box with classical inputs and outputs. Quantum steering has been receiving lot of attention recently due to the fact that it allows for entanglement detection which is more robust to noise and experimental imperfections than Bell nonlocality \cite{WJD07,Quintino15}. Moreover, quantum steering was shown to be useful for one-sided device independent quantum key distribution \cite{1sided} and randomness certification \cite{RandSteering}. Several experimental groups have recently observed steering, including in continuous-variable systems \cite{exp steering0,exp steering1}, using Bell local states \cite{exp steering2}, using inefficient detectors \cite{exp steering3,exp steering4,exp steering5}, asymmetric states \cite{exp steering6}, and multipartite systems \cite{multi steering1,multi steering2,multi steering3}.

The main result of our paper is a general and optimal method to quantify the amount of local or global randomness that can be certified from a single measurement in a steering experiment. We use this method to show that local randomness can be certified provided that the detectors used have efficiency higher than $50\%$. Our method can be seen as the analogue of the approach of \cite{NPS14,BSS14} from the fully-device-independent scenario applied to the steering scenario. We compare the results obtained there to those obtained here, in terms of the amount of randomness that can be obtained by measuring systems subjected to white noise, and find substantial benefits can be obtained in the present setting. As a by-product, we also show that the amount of randomness certified in Ref. \cite{RandSteering} from the two-qubit Werner state is optimal. 

We furthermore show that the results can be easily extended beyond the steering scenario, to the prepare-and-measure scenario, where the state is also trusted, so that only Alice's measuring device is untrusted. In this case we show that even noisy states can perform very well for randomness certification.

Finally, we give a method to find the best measurements which obtain the most randomness from any fixed state. Using insight from this method, we show analytically that all pure partially entangled states lead to maximal randomness certification using only two fixed measurements.

There are several motivations to quantify the amount of randomness in the steering scenario. From a fundamental point of view, it is important to understand how much randomness can be maintained if we give up partial information about the specific description of the systems \cite{RandSteering,semi-dir,semi-dir2}. From a practical point of view, the amount of randomness obtained in the steering scenario gives an upper bound to what Alice and Bob would obtain in a fully device-independent setting, regardless of the number of measurements Bob would apply. Furthermore, it is a scenario that appears naturally in some asymmetric applications. For instance the present results give a way of quantifying the amount of randomness in remote untrusted stations. This is relevant, for instance, when the provider of a quantum-random-number generator wants to remotely check if the devices they provided are still functioning properly.

\section{Steering and randomness}
\begin{figure}[t]
\begin{center}
  \includegraphics[width=\columnwidth]{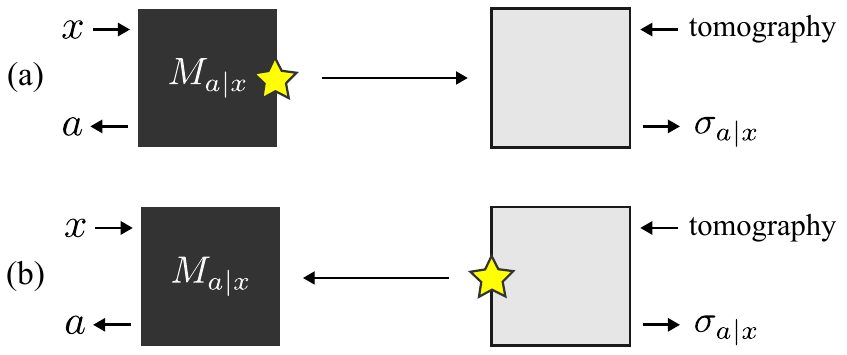}
  \end{center}
  \caption{Setup for randomness certification in the quantum steering and prepare-and-measure scenarios.
    (a) Steering scenario: Alice and Bob measure an unknown bipartite system delivered by an untrusted source. Alice treats her measurement device as a black box  with inputs $x \in \{0,\ldots,m_\rA-1\}$ and outputs $a \in \{0,\ldots,n_\rA-1\}$ and Bob performs tomography on his subsystem. (b) Prepare-and-measure scenario: similar to the previous scenario, but now Bob holds the source and then knows the bipartite state $\rho_{\rA\rB}$.}
  \label{fig:setup}
\end{figure}

The scenario we treat in this work is the following~\cite{WJD07}: two parties, Alice and Bob, are located in distant laboratories and receive a bipartite system from a source. One of the two parties, say Alice, does not trust her measuring devices, which are treated as ``black boxes''. She can, nevertheless, choose which measurement to perform, which she labels by $x \in \{0,\ldots,m_\rA-1\}$, each of which provides an outcomes, which she labels $a\in \{0,\ldots,n_\rA-1\}$. The other party, Bob, has complete knowledge of his device, which allows him to perform quantum state tomography on his part of the system, and thus to obtain a complete description of his subsystem (see Fig.~\ref{fig:setup} (a)). The states reconstructed by Bob will usually depend on Alice's input and output as $\rho_{a|x}=\Tr_\rA[(M_{a|x} \otimes \identity_\rB)\rho_{\rA\rB}]/P(a|x)$, where $\rho_{AB}$ is the unknown state shared with Alice, $P(a|x)$ is the probability that Alice observes outcome $a$ given she chose $x$, and $M_{a|x}$ is the corresponding (unknown) element of Alice's measurement. The set of unnormalized states $\sigma_{a|x} = \Tr_\rA[(M_{a|x} \otimes \identity_\rB)\rho_{\rA\rB}]=\rho_{a|x}P(a|x)$ is called an \textit{assemblage} and can be completely determined by Bob through tomographic measurements.

As noticed in \cite{WJD07}, Bob can determine if $\rho_{\rA\rB}$ is entangled by looking at the form of the assemblage $\{\sigma_{a|x}\}_{a,x}$. This is because separable states can only lead to assemblages with the specific form
\begin{equation}\label{eq:lhs}
\sigma_{a|x} = \sum_{\lambda} q(\lambda) P(a|x,\lambda) \sigma_{\lambda},
\end{equation}
where $\lambda$ is a hidden variable distributed according to $q(\lambda)$, which determines both Alice's response $P(a|x,\lambda)$, and the states sent to Bob, $\sigma_\lambda$.  Assemblages of this form are said to have a Local Hidden State (LHS) model. Any assemblage which does not have this form can be detected through the violation of a steering inequality \cite{CJWR09} (similar to a Bell inequality or an entanglement witness) or a simple semi-definite program \cite{Pusey}.

It turns out that the confirmation of steering not only guarantees that the shared state is entangled, but also that Alice is performing incompatible measurements \cite{steering vs JM,steering vs JM2}. It is thus very intuitive to expect a relation between steering and randomness: first, the correlations (entanglement) shared between Alice and Bob allows Bob to certify steering, and consequently the incompatibility of Alice's measurements. Second, since Alice's measurements are incompatible not all the outcomes she receives are predictable, and thus random.

\section{Local randomness certification}\label{sect:pguess}
In order to certify the local randomness of Alice's outcomes we work in the adversarial scenario, where a potential eavesdropper, Eve, wants to predict them. This framework is relevant for cryptographic tasks, namely 1SDIQKD.
In the most general case, we do not make any assumption on Alice's measurement device, so that it could even have been provided by Eve. We also consider that the state $\rho_{\rA\rB}$ is the reduced state of a tripartite entangled state $\rho_{\rA\rB\rE}$ shared by Alice, Bob and Eve, \ie $\rho_{\rA\rB} = \Tr_\rE [\rho_{\rA\rB\rE}]$.
Hence, by applying measurements to her subsystem Eve can in principle obtain information about Alice's outcome.

In this section we will focus on the case where Alice and Bob want to extract randomness from the outcomes of a single given measurement of Alice, let us say $x^* \in \{0,\ldots,m_A-1\}$. The motivation for considering this case is that it is the relevant one from the perspective of 1SDIQKD. We assume that the runs of the experiment are independent and identically distributed with respect to Eve's strategy\footnote{We note that once independence is assumed, it is without loss of generality to assume the pairs identical.}. We consider the case where Eve also knows from which measurement $x^*$ Alice is going to extract randomness, so she can optimise her attack to obtain information about this measurement setting.
The figure of merit we use to evaluate the amount of randomness in Alice's outcomes is the probability that Eve can correctly guess the outcome $a$ of the measurement $x^*$ of Alice. This quantity, denoted by $P_{\guess}(x^*)$, is given by the probability that Eve's guess $e$ is equal to the outcome $a$ that Alice obtained, whenever Alice performs the specific measurement $x=x^*$:
\begin{equation}\label{eq:pguess}
P_{\guess}(x^*) = \sum_e P_\rA(a=e|x^*) \, P_\rE(e|a=e,x=x^*) .
\end{equation}
Applying Bayes theorem, this is equivalent to $P_{\guess}(x^*) = \sum_e P_{\rA\rE}(a=e,e|x=x^*)$, i.e. equal to the joint probability that Alice and Eve give the same outcome whenever Alice measures $x=x^*$.
Randomness is certified whenever the guessing probability is strictly less than 1, in which case Eve can not predict Alice's outcome with certainty.

After Alice and Eve have applied their measurements the assemblage prepared will be
\begin{equation}
\sigma_{a|x}^e = \Tr_{\rA\rE}[(M_{a|x} \otimes \identity_{\rB} \otimes M_e) \, \rho_{\rA\rB\rE}],
\end{equation}
where $M_e$ is the element of Eve's (optimal) measurement which yields outcome $e \in \{0,\ldots,n_\rA-1\}$.
However, since Alice and Bob do not have access to Eve's outcomes the assemblage they will reconstruct will be given by
\begin{equation}
\sigma_{a|x}^\obs=\sum_e\sigma_{a|x}^e.
\end{equation}

In order to compute the optimal strategy for Eve we need to maximise her guessing probability (for a given input $x^*$ of Alice), over all strategies. Naively, this would appear to constitute optimising the triple $\{\rho_{\rA\rB\rE}, M_{a|x}, M_e\}$, of state, measurements for Alice, and measurement for Eve, a non-linear optimisation problem. However, just as in the device-independent case \cite{NPS14,BSS14}, we can instead replace this by an equivalent linear optimisation over all physical assemblages $\{\sigma_{a|x}^e\}_{a,e,x}$ that are compatible with the no-signalling principle and the observed assemblage $\{\sigma_{a|x}^\obs\}_{a,x}$.
More precisely, the maximisation problem can be formulated as the following semidefinite programme (SDP) \cite{boyd}:
\begin{align} \label{eq:sdp}%
P_\guess(x^*)=\max_{\{\sigma_{a|x}^e\}_{a,e,x}} &\sum_e \Tr[\sigma_{a=e|x^*}^e] \\
\mathrm{s.t.} \quad &\sum_e \sigma_{a|x}^e = \sigma_{a|x}^\obs  &\forall \, a,x \nonumber \\
&\sum_a \sigma_{a|x}^e = \sum_a \sigma_{a|x'}^e &\forall \, e, \, x \neq x' \nonumber \\
&\sigma_{a|x}^e \succeq 0 &\forall a,x,e. \nonumber
\end{align}
In the objective function we used $P_\rE(e)P_\rA(a|x,e) = P(ae|x)=\Tr[\sigma_{a|x}^e]$ to re-express $P_{\guess}(x^*)$. The first constraint assures that the decomposition for Eve is compatible with the assemblage Alice and Bob observe. The second constraint is the non-signalling condition -- i.e. Alice cannot signal to Bob and to Eve. The last one is the requirement for every $\sigma_{a|x}^e$ to be a valid (unnormalized) quantum state. We defer to the appendix the full proof that this optimisation problem is equivalent to optimising over states and measurements, which follows from the Gisin-Hughston-Jozsa-Wootters (GHJW) theorem \cite{HJW} (which shows that all bipartite no-signalling assemblages have quantum realisations), combined with the fact that Eve, making only one measurement, also cannot signal.

Notice that the SDP (\ref{eq:sdp}) can be seen as the steering analogue of the SDP provided in \cite{NPS14,BSS14} which bounds the amount of randomness given an observed nonlocal probability distribution $P_\obs(ab|xy)$. As mentioned before, the SDP (\ref{eq:sdp}) provides an upper bound on the amount of randomness (\ie a lower bound on the $P_{\guess}$) that can be found using the SDP of \cite{NPS14,BSS14}. This follows because (\ref{eq:sdp}) does not allow Eve to attack the measurements of Bob. Thus, our SDP bounds the maximal amount of randomness that could be obtained if Bob were to perform any number of measurements (that Eve can attack) and compute the randomness based on the obtained probability distribution.
The number of random bits is quantified by the min-entropy $H_{\min}(A|X) = - \log_2 P^{*}_{\guess}(x^*)$, where $P^{*}_{\guess}(x^*)$ is the result of the maximization (\ref{eq:sdp}).

\begin{figure}
  \begin{center}
    \includegraphics[width=\columnwidth]{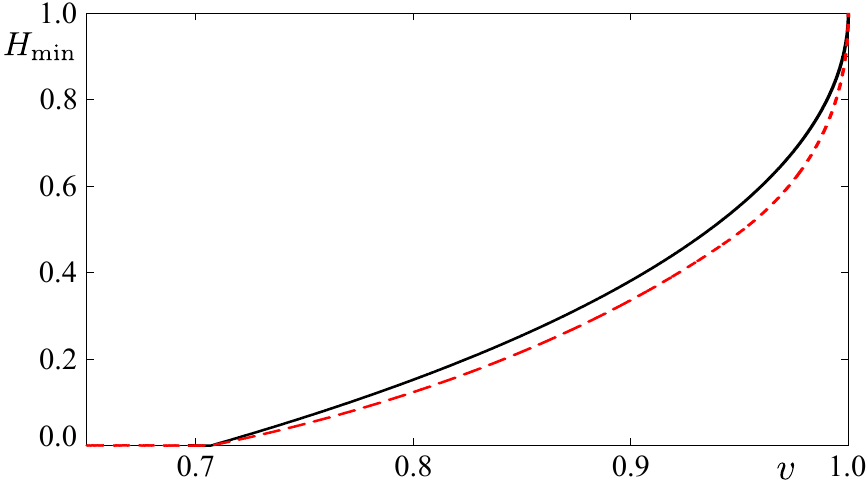}
  \end{center}
  \caption{\label{fig:plot r vs v}Random bits certified $H_{\min}$ versus the visibility $v$ of the two-qubit Werner state. We compare the randomness obtained with our method in the steering scenario (solid line) with the fully-device-independent case as in \cite{NPS14} (dashed line).}
  \begin{center}
    \includegraphics[width=\columnwidth]{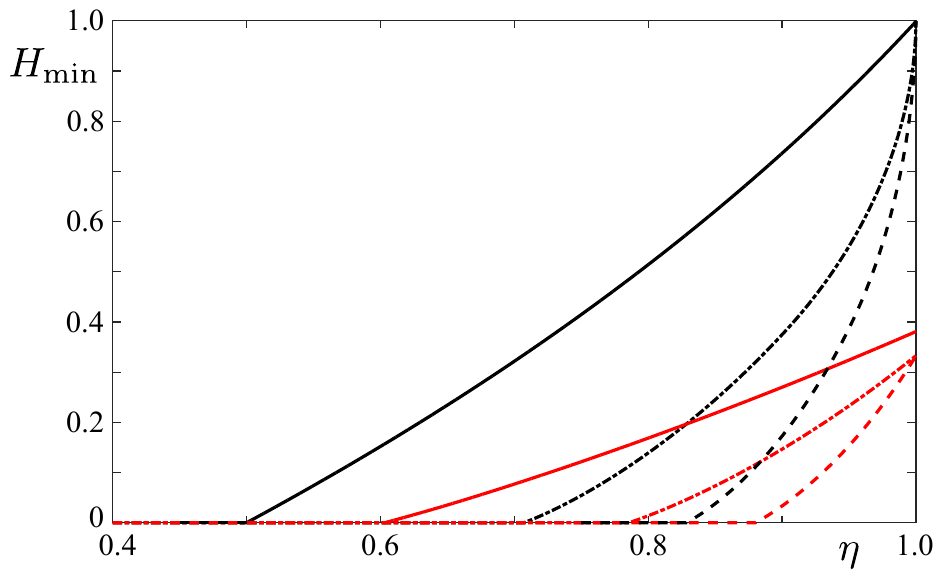}
  \end{center}
  \caption{\label{fig:HminVSetav1}Random bits certified $H_{\min}$ versus the detection efficiency $\eta$ for the two-qubit Werner state. Black lines: $v = 1$; Red lines: $v = 0.9$. Solid lines: our steering method; Dot-dashed lines: DI method in the case where Bob's detection efficiency is 1; Dashed lines: DI method where both Alice and Bob's detectors have efficiency $\eta$.}
\end{figure}

In Fig.~\ref{fig:plot r vs v} we plot the amount of randomness certified in the case that Alice applies two mutually unbiased Pauli spin measurements on a two-qubit Werner state $\rho_{\rA\rB}=v\ket{\Phi_+}\bra{\Phi_+}+(1-v)\identity/4$, where $\ket{\Phi_+}=(\ket{00}+\ket{11})/\sqrt{2}$, and compare it with the amount of randomness obtained in the case Bob also treats his measuring device as a black box (i.e. the fully device-independent case). In both cases randomness can be certified as long as $v > 1/\sqrt{2}$, which is the critical amount of noise for demonstrating either steering or nonlocality with only two measurements \cite{CavJonWis2009}. All numerical SDP calculations were performed using the {\sc cvx} package for {\sc matlab} \cite{cvx}, along with the library {\sc qetlab} \cite{qetlab}.

In Fig. \ref{fig:HminVSetav1} we also compute the amount of randomness that can be obtained by measuring the same spin measurements with detection efficiency $\eta$ (for visibility $v = 1$ and $v = 0.9$), again comparing to the case where Bob treats his measuring device as a black box. That is, (for steering) instead of ideal measurements, with elements $M_{a|x}$, we consider inefficient measurements $M^{(\eta)}_{a|x}$, with one additional outcome $a = \emptyset$, given by
\begin{equation}
M^{(\eta)}_{a|x} = \Bigg\{\begin{array}{cc}
\eta M_{a|x}, & \qquad a \neq \emptyset \\
(1-\eta) \openone, & \qquad a = \emptyset
\end{array}
\end{equation}
(the measurements of Bob are similarly made inefficient in the nonlocality scenario).

In this case, two comparisons are made: (i) the case where Bob's detection efficiency is 1; and (ii) where Bob also has detection efficiency $\eta$. As one can see, for $v = 1$ in the steering scenario randomness can be certified whenever the detection efficiency is higher than $50\%$,  matching the threshold below which no randomness can be obtained \cite{attack}. Moreover, we see that due to the much larger detection efficiencies needed to violate the CHSH inequality ($82.8\%$) and for the DI case where Bob's measuring device is perfectly efficient ($70.7\%$), the steering scenario offers a significant advantage when using the maximally entangled state over the nonlocality scenario, for the entire range of visibility which is experimentally significant (i.e. for $v=0.9$ and above).

\begin{figure}
\begin{center}
  \includegraphics[width=\columnwidth]{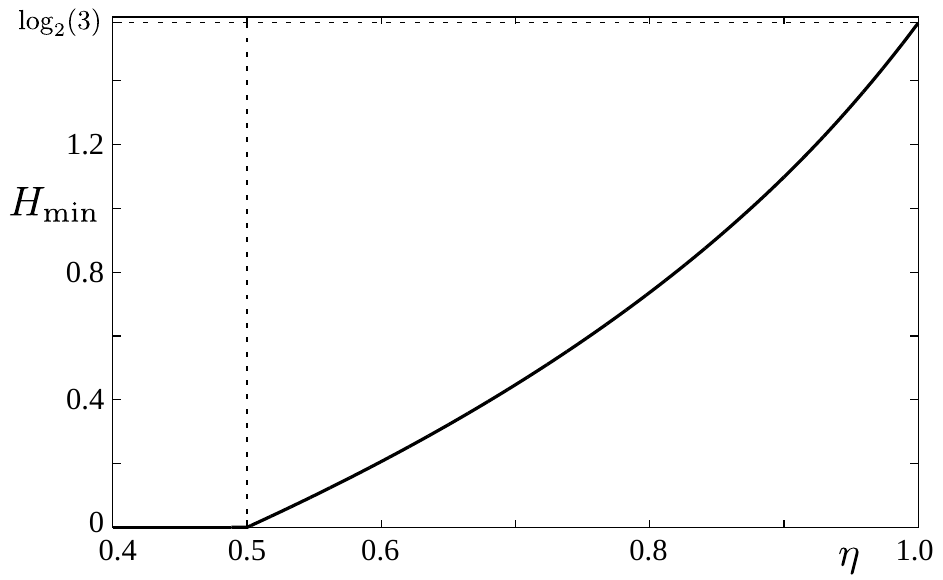}
\end{center}
\caption{\label{fig:r vs eta qutrit}Random bits certified $H_{\min}$ versus the detection efficiency $\eta$ for the two-qutrit maximally entangled state $\ket{\Phi_+^{(3)}}=(\ket{00}+\ket{11}+\ket{22})/\sqrt{3}$.}
\end{figure}

Finally, in Fig. \ref{fig:r vs eta qutrit} we plot the number of random bits certified in the case that Alice performs measurements in four mutually unbiased bases on her half of the entangled two-qutrit state $(\ket{00}+\ket{11}+\ket{22})/\sqrt{3}$ in the presence of losses. Again, we see that whenever the detection efficiency is above $50\%$ Alice is able to certify local randomness. Moreover, for efficiency $\eta = 1$ she certifies $H_\mathrm{min} = \log_2 3$ bits of randomness.

\section{Global randomness certification}
In the steering scenario one can also consider global randomness extraction from both the untrusted and trusted devices. Indeed, even though Bob trusts his devices, and knows which measurement he performs, there is still an optimal state that Eve can distribute which allows her to predict the outcome of Bob's measurement. This is because although Eve is not able to change the measurements performed by Bob, nor his reduced state, she still has additional classical side information that she can use to help her in guessing the result of Bob (since she holds the source). 
\begin{figure}[t]\begin{center}
  \includegraphics[width=\columnwidth]{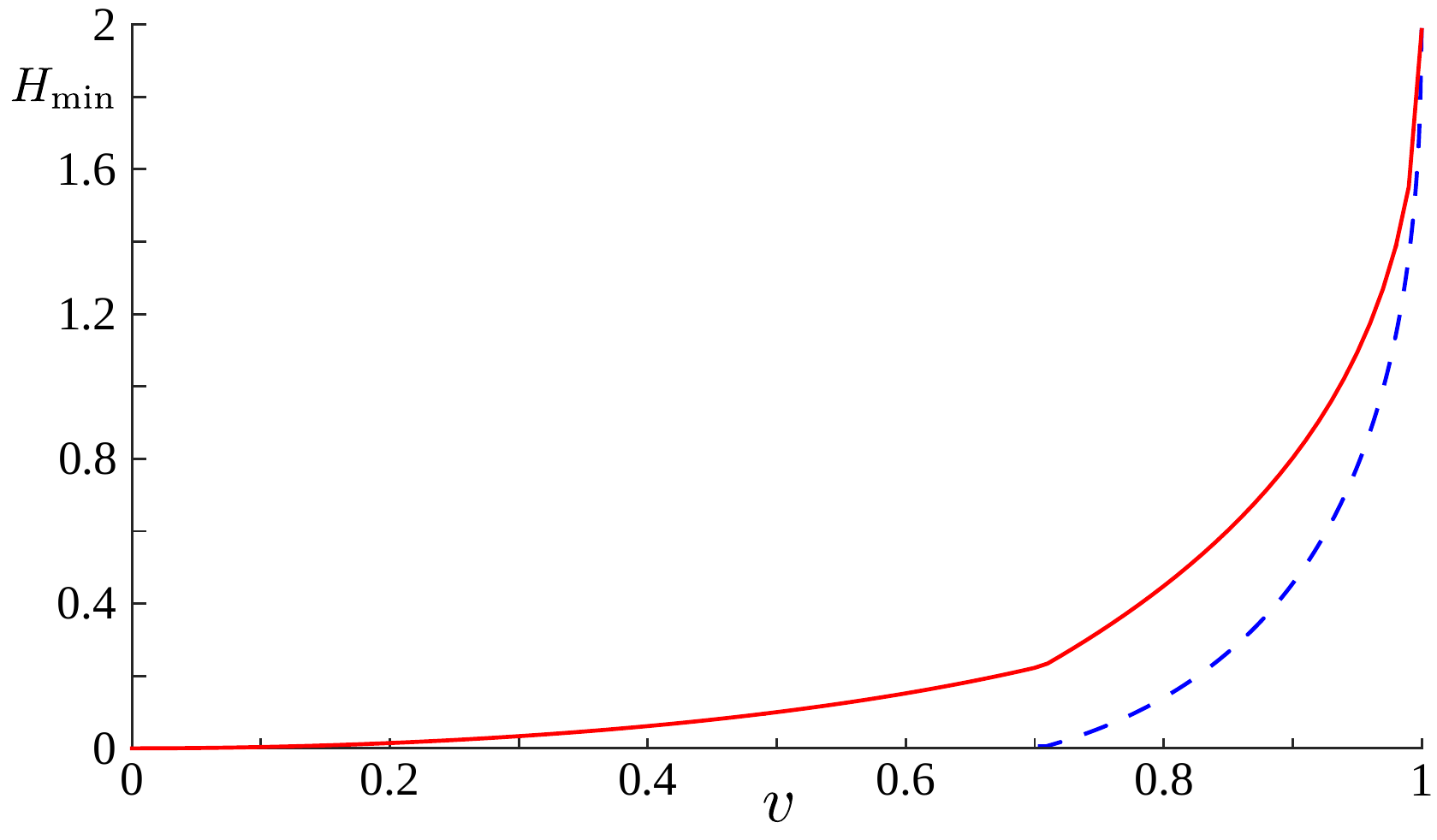}\end{center}
  \caption{Global randomness obtained by measuring a two-qubit Werner state (with noise $v$), with $X$ and $Z$ measurements for Alice, and $X$ measurement for Bob, computed using Eq.~(\ref{e:global sdp}) (solid curve). As a matter of comparison we also plot the amount of global randomness obtained in the device-independent scenario, using the methods of Refs.~\cite{NPS14,BSS14} (dashed curve). }
  \label{fig:global}
\end{figure}

Consider that, additionally to Alice's measurement $x=x^*$, Eve wants to guess the outcomes of a measurement $M_b$ performed by Bob. Eve now has a pair of guesses $(e,e')$, which will be her guess for the pair $(a,b)$. She will thus perform a measurement with elements $M_{ee'}$ on her share of the state, which after Alice also measures will lead to the assemblage for Bob $\sigma_{a|x}^{ee'} = \tr_{\rA\rE}[(M_{a|x}\otimes \openone_\rB \otimes M_{ee'})\rho_{\rA\rB\rE}]$. Similarly to the case of local randomness, the global guessing probability $P_\mathrm{g}$ can straightforwardly be shown to be the solution to the following SDP
\begin{align} \label{e:global sdp}
P_\guess(x^*) = \max& \sum_{e e'} \Tr [ M_{b=e'} \sigma_{a=e|x^*}^{e e'} ]   \\
  \mathrm{s.t.}& \sum_{ee'} \sigma_{a|x}^{ee'} = \sigma_{a|x}^\mathrm{obs}, &\forall a,x \nonumber  \\
 &\sum_a \sigma_{a|x}^{ee'} = \sum_a \sigma_{a|x'}^{ee'}, &\forall x\neq x', a,e,e' \nonumber \\
 &\sigma_{a|x}^{ee'} \succeq 0, &\forall a,x,e,e' \nonumber
\end{align}
We again require consistency with the observed assemblage $\sigma_{a|x}^\mathrm{obs}$, and demand positivity and no-signalling.

We computed the global randomness which can be certified without losses assuming $X$ and $Z$ measurements for Alice, and an $X$ measurement for Bob, on two-qubit Werner states. The results can be seen in Fig.~\ref{fig:global}, alongside the corresponding curve calculated using the method of Refs. \cite{NPS14,BSS14} for the nonlocality scenario. As a result, we observe that the lower bound on the amount of global randomness that can be extracted in the steering scenario presented in Ref. \cite{RandSteering} is tight.

\section{Prepare-and-measure scenario}
Up to now we have considered the steering scenario, where Alice and Bob receive an unknown state $\rho_{\rA\rB}$ from an untrusted source. It turns out that the results on local randomness straightforwardly apply to the case where Bob prepares a known state and sends half of it to Alice (see Fig.~\ref{fig:setup} (b)). In this case, since the global state $\rho_{\rA\rB}$ is known, the assemblages reconstructed by Bob have to come from unknown measurements on this state, \ie $\sigma_{a|x}=\sum_e\Tr_\rA[(M^e_{a|x}\otimes\openone_\rB )\rho_{\rA\rB}]$. Thus the SDP (\ref{eq:sdp}) can be replaced by
\begin{align} \label{eq:3sdp}
P_{\guess}(x^*)=\max & \sum_e \Tr[(M^e_{a=e|x^*}\otimes\openone_\rB) \rho_{\rA\rB}] \\
\mathrm{s.t.} &\sum_e \Tr_\rA[(M^e_{a|x}\otimes\openone_\rB) \rho_{\rA\rB}] = \sigma_{a|x}^\obs, \,\,\,\forall \, a,x \nonumber \\
& \sum_a M_{a|x}^e = \sum_a M_{a|x'}^e \qquad\qquad \forall x'\neq x, e  \nonumber \\
&\sum_{a,e} M_{a|x}^e = \openone \,\,\qquad\qquad\qquad\qquad\qquad\forall x  \nonumber \\
&M_{a|x}^e \succeq 0 \,\,\qquad\qquad\qquad\qquad\qquad\forall a,x,e \nonumber
\end{align}
This SDP can be understood as the maximisation of Eve's guessing probability over all possible POVM measurements (where the outcome $e$ goes to Eve and the outcome $a$ goes to Alice), with Eve oblivious of $x$, that can be applied to the state $\rho_{\rA\rB}$, given the observation of the assemblage $\{\sigma_{a|x}^\obs\}_{a,x}$. A derivation of this SDP can be found in \ref{a:PM SDP}. We note that this scenario can also be thought of as the `time-like steering' scenario introduced in Ref.~\cite{Pusey2015}.

We used the above program to calculate the amount of randomness that can be obtained from the two qubit Werner state, and from the isotropic two-qutrit state $\rho_{\rA\rB}=v\ket{\Phi_+^{(3)}}\bra{\Phi_+^{(3)}}+(1-v)\openone/9$, where $\ket{\Phi_+^{(3)}}=(\ket{00}+\ket{11} + \ket{22})/\sqrt{3}$. In both cases we consider that Alice performs two mutually unbiased measurements (Pauli $X$ and $Z$ for qubits, and their generalisation for qutrits). 

For the case of no-losses, we observe that the amount of randomness that can be extracted is \textit{independent of the visibility $v$}, and equal to 1 bit and 1 trit = $\log_2(3)$ bits respectively\footnote{More precisely, for all $v \geq 0.05$ we observed numerically that $P_g \leq 0.339$.} This coincides with the amount which is obtained in the steering scenario for $v = 1$, i.e. the ideal case. This demonstrates that if knowledge of the state is assumed, then the lack of visibility cannot be used by Eve to guess the outcomes of Alice's measurements.

Turning to the case of losses, consistent with the above, we observe that, independent of the visibility, the dependence of the randomness on the loss coincides with that found in the steering scenario for perfect visibility. That is, the solid black curves in Figs.~\ref{fig:HminVSetav1} and \ref{fig:r vs eta qutrit} are obtained, for any fixed value of the visibility $v$. 

This shows that the prepare-and-measure scenario greatly improves over the steering scenario when considering lack of visibility (i.e. noise) on the state.

\section{Improving the randomness extraction}
\begin{figure}[t]
  \begin{center}
    \includegraphics[width=0.9\columnwidth]{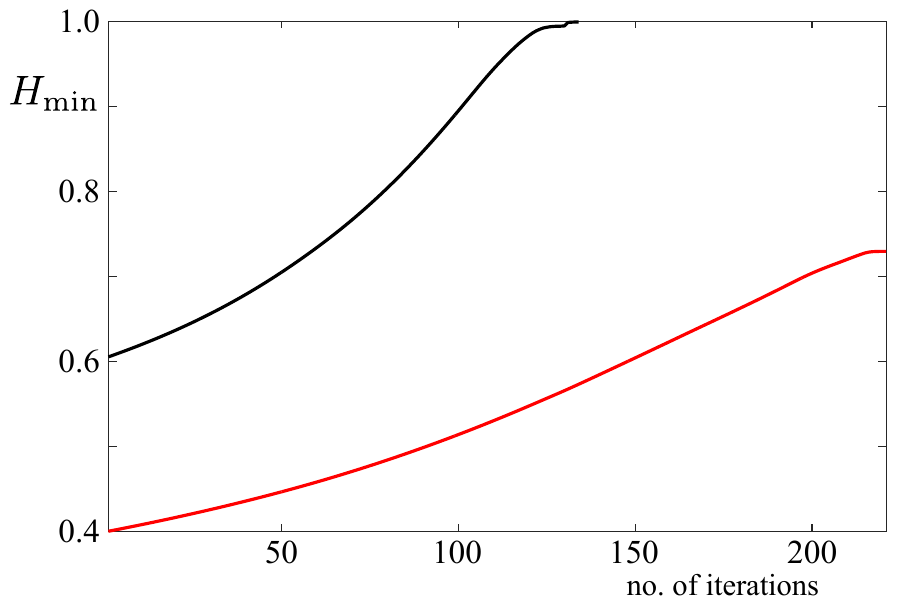}
  \end{center}
  \caption{\label{fig:plot r vs steps}Plot of the random bits certified versus the number of steps of the see-saw iteration for a two-qubit partially entangled state $\ket{\psi} = \cos \theta \ket{00} + \sin \theta \ket{11}$ with $\theta = \pi/7$ and starting with random measurements with $\eta = 1$ (black curve) and $\eta = 0.9$ (red curve).} 
\end{figure}

The SDP (\ref{eq:sdp}) provides a way of quantifying the randomness in Alice's outcomes given the observation of a given assemblage. A natural question is, given a fixed state distributed between Alice and Bob and a fixed number of measurements for Alice, what is the best scheme they can implement (i.e. the best choice of measurements) which allows for the certification of the most randomness.

Here we propose a numerical see-saw method that, starting from an initial amount of certified randomness, seeks for measurement schemes that lead to higher randomness certification. We focus on the case of local randomness. A similar scheme can also be implemented for global randomness. 

Every SDP has a dual program, also an SDP, that can be obtained through the theory of Lagrange multipliers \cite{boyd}. The dual of (\ref{eq:sdp}) is equivalent to
\begin{align} \label{eq:dual}
  \min_{\{F_{a|x}\}_{a,x}} \quad&\sum_{a,x} \Tr[F_{a|x} \, \sigma_{a|x}^\obs] \\
  \mathrm{s.t.} \quad&\Tr[\sigma_{a'|x^*}] \leq \sum_{a,x} \Tr[F_{a|x} \sigma_{a|x}]  \qquad \forall \, a', \sigma_{a|x} \nonumber
\end{align}
where in the constraint, $\forall \sigma_{a|x}$ should be understood as for all non-signalling assemblages, i.e. those satisfying $\sum_a \sigma_{a|x} = \sum_a \sigma_{a|x'}$ for all $x'\neq x$ \footnote{As written, this problem is not in the form of an SDP. In \ref{a:dual} we derive the dual SDP and show its equivalence to (\ref{eq:dual}), which is easier to interpret.}. Since strong duality holds, the optimal value of this optimisation problem is equal to the optimal value of (\ref{eq:sdp}), i.e. $P^{*}_{\guess}(x^*) = \sum_{a,x} \Tr(F^*_{a|x} \, \sigma_{a|x}^\obs)$.  Moreover, it outputs the coefficients $F^*_{a|x}$ of the optimal steering inequality that gives the tight upper bound on $P^{*}_{\guess}(x^*)$.

Once we have solved the dual problem (\ref{eq:dual})  we can run a second SDP that optimizes the violation of the steering inequality $\sum_{a,x} \Tr(F^*_{a|x} \, \sigma_{a|x})$ over Alice's measurements $\{M_{a|x}\}_{ax}$:
\begin{align}\label{eq:sdp-meas}
\min_{\{M_{a|x}\}_{ax}} \quad &\sum_{ax} \Tr[(M_{a|x} \otimes F_{a|x}) \rho_{AB}]  \\
\mathrm{s.t.} \quad &\sum_a M_{a|x} = \openone  &\forall \, x \nonumber \\
&M_{a|x} \succeq 0 &\forall a,x\nonumber
\end{align}
The solution of this optimisation problem provides the measurements for Alice that allow for the certification of the most randomness using the steering inequality provided by the first SDP.

At this point, one can perform a see-saw iteration of the two SDPs in order to obtain the maximal randomness that can be certified from a given state, along with the optimal steering inequality and measurements $M_{a|x}$.
For every given initial state, the SDP (\ref{eq:sdp}) (and its dual (\ref{eq:dual})) gives the best inequality to certify randomness from an assemblage, while the SDP (\ref{eq:sdp-meas}) gives the best set of measurements -- and therefore the best assemblage -- for a given steering inequality.

In Fig.~\ref{fig:plot r vs steps} we plot the result of this see-saw iteration, starting from two randomly chosen projective measurements, for $\eta = 1$ and $\eta= 0.9$, for the two-qubit partially entangled state $\ket{\psi} = \cos \theta \ket{00} + \sin \theta \ket{11}$. When there are no losses, one bit of randomness is already known to be possible from any partially entangled state in the fully device-independent scenario \cite{AciMasPir12}. Since this scenario is more demanding, it implies one bit can also be obtained from any partially entangled state of two qubits in the steering scenario. If the method works it should be able to reproduce this result. As can be seen, 1 bit of randomness is indeed found, thus demonstrating the utility of the method.

Further exploration showed numerically that the measurements which achieve 1 bit of randomness from any partially entangled state can always be taken to be $X$ and $Z$ measurements for Alice (with the randomness obtained from the $X$ measurement)\footnote{We do not present the form of the optimal steering inequalities for partially entangled states, since we did not find any general structure which makes knowing their form useful.}.

In the appendix we show that this numerical evidence can be turned into an analytic construction, which proves that 1 bit can be obtained from any partially entangled state of two qubits (which is notably completely different to the approach used in \cite{AciMasPir12} for nonlocality). Moreover, the construction generalises to qudits in a straightforward manner, showing that 1 dit of randomness can be obtained by performing two generalised Pauli measurements on any Schmidt-rank $d$ state. This is contrary to the fully device-independent case, where it is only known how to extract 1 bit from pure partially entangled states.

\section{Conclusions}
We have presented a method that certifies the optimal amount of local or global randomness that can be extracted in a steering experiment. We also considered the case where the source is trusted (prepare-and-measure scenario).
Our method relies on optimisation techniques that quantify the amount of certified randomness and provide the optimal steering inequality for randomness certification.
Applying this method to realistic implementations - \ie in presence of noise and losses - we have shown that a detection efficiency above $50\%$ is sufficient to have reliable local randomness certification in the steering scenario.
This result is also valid for device-independent (DI) randomness certification and, in general, in scenarios with lower levels of trust.

Finally, we have introduced a method which produces, for any given initial state, the optimal measurements which in turn give the optimal assemblage from which maximal randomness can be certified. Using this method as a starting point, we have shown analytically that 1 dit of randomness can be obtained from any pure entangled Schmidt-rank $d$ state. 

Since local randomness certification is of fundamental importance for 1SDIQKD and DIQKD, the results presented here have natural applications in cryptographic protocols.

\section*{Acknowledgements}
We thank R. Rabelo for discussions on randomness in an early stage of this project. This work was supported by the Beatriu de Pin\'os fellowship (BP-DGR 2013),the Marie Curie COFUND action through the ICFOnest program, the ERC CoG QITBOX, the ERC AdG NLST, the EU project SIQS, the Spanish project FOQUS, the Generalitat de Catalunya (SGR875) and the John Templeton Foundation. E.~P. acknowledges the Max Planck Institute for Quantum Optics for hospitality.

\bibliographystyle{iopart-num}

\begin{thebibliography}{}

\bibitem{QRNG} J. G. Rarity, P. C. M. Owens, P. R. Tapster, J. Mod. Optic, {\bf 41}(12), 2435 (1994).
\bibitem{QKD_RMP} V. Scarani, H. Bechmann-Pasquinucci, N. J. Cerf, M. Dusek, N. Lutkenhaus, M. Peev, Rev. Mod. Phys. {\bf 81}, 1301 (2009).
\bibitem{Bell64} J. S. Bell, Physics (College. Park. Md). \textbf{1}, 195 (1964).
\bibitem{Bell review} N. Brunner, D. Cavalcanti, S. Pironio, V. Scarani and S. Wehner, 
Rev. Mod. Phys. \textbf{86}, 419 (2014).
\bibitem{Colbeck} R. Colbeck, PhD thesis, University of Cambridge (2006), arXiv:0911.3814 (2009).
\bibitem{PAMBMMOHLMM10} S. Pironio \emph{et al.}, Nature {\bf 464}, 1021 (2010).
\bibitem{hack1} L. Lydersen et al., 
Nat. Phot. \textbf{4}, 686 (2010).
\bibitem{hack2} I. Gerhardt et al., 
Nature Comm. \textbf{2}, 349 (2011).
\bibitem{hack3} I. Gerhardt, Q. Liu, A. Lamas-Linares, J. Skaar, V. Scarani, V. Makarov, C. Kurtsiefer, 
Phys. Rev. Lett. {\bf107}, 170404 (2011).
\bibitem{Schr} E. Schrodinger, Proc. Camb. Phil. Soc. {\bf 31}, 555 (1935).
\bibitem{WJD07} H. M. Wiseman, S. J. Jones, and A. C. Doherty, Phys. Rev. Lett. {\bf 98}, 140402 (2007).
\bibitem{Quintino15} M. T. Quintino, T. V\'ertesi, D. Cavalcanti, R. Augusiak, M. Demianowicz, A. Ac\'in, N. Brunner, Phys. Rev. A \textbf{92}, 032107 (2015).
\bibitem{1sided} C. Branciard, E. G. Cavalcanti, S. P. Walborn, V. Scarani, H. M. Wiseman, 
Phys. Rev. A \textbf{85}, 010301(R) (2012).
\bibitem{RandSteering} Y. Z. Law, L. P. Thinh, J. D. Bancal, V. Scarani, 
J. Phys. A: Math. Theor. {\bf 47}, 424028 (2014).
\bibitem{exp steering0} Z. Y. Ou, S. F. Pereira, H. J. Kimble, and K. C. Peng,
Phys. Rev. Lett. {\bf 68}, 3663 (1992).
\bibitem{exp steering1} W. P. Bowen, R. Schnabel, P. K. Lam, and T. C. Ralph,
Phys. Rev. Lett. {\bf 90}, 043601 (2003).
\bibitem{exp steering2} D. J. Saunders, S. J. Jones, H. M. Wiseman and G. J. Pryde, 
Nat. Phys. {\bf 6}, 845 (2010).
\bibitem{exp steering3} D.-H. Smith \emph{et al.}, 
Nat. Commun. {\bf 3}, 625 (2012).
\bibitem{exp steering4} A.~J. Bennet \emph{et al.}, 
    Phys. Rev. X {\bf 2}, 031003 (2012).
\bibitem{exp steering5} B. Wittmann \emph{et al.}, 
New J. Phys. {\bf 14}, 053030 (2012).
\bibitem{exp steering6} V. H\"andchen \emph{et al.}, 
Nat. Phot. {\bf 6}, 598 (2012).
\bibitem{multi steering1} S. Armstrong, M. Wang, R. Y. Teh, Q. Gong, Q. He, J. Janousek, H.-A. Bachor, M. D. Reid, and P. K. Lam, Nat. Phys. \textbf{11}, 167-172 (2015).
\bibitem{multi steering2} D. Cavalcanti, P. Skrzypczyk, G. H. Aguilar, R. V. Nery, P. H. Souto Ribeiro, S. P. Walborn, Nat. Commun. \textbf{6}, 7941 (2015).
\bibitem{multi steering3} C.-M. Li, K. Chen, Y.-N. Chen, Q. Zhang, Y.-A. Chen, J.-W. Pan, Phys. Rev. Lett. \textbf{115}, 010402 (2015).
\bibitem{NPS14} O. Nieto-Silleras, S. Pironio, J. Silman, New J. Phys. {\bf 16}, 013035 (2014).
\bibitem{BSS14} J. D. Bancal, L. Sheridan, V. Scarani, New J. Phys. {\bf 16}, 033011 (2014).
\bibitem{semi-dir} J. Bowles, M. T. Quintino, and N. Brunner,
Phys. Rev. Lett. {\bf 112}, 140407 (2014).
\bibitem{semi-dir2}
H.-W. Li, M. Pawlowski, Z.-Q. Yin, G.-C. Guo, Z.-F. Han, Phys. Rev. A \textbf{85}, 052308 2012.
\bibitem{CJWR09} E. G. Cavalcanti, S. J. Jones, H. M. Wiseman, and M. D. Reid, Phys. Rev. A {\bf 80}, 032112 (2009).
\bibitem{Pusey} M. F. Pusey, Phys. Rev. A \textbf{88}, 032313 (2013).
\bibitem{steering vs JM} M. T. Quintino, T. V\'ertesi, and N. Brunner,
Phys. Rev. Lett. {\bf 113}, 160402 (2014).
\bibitem{steering vs JM2} R. Uola, T. Moroder, and O. G\"{u}hne,
Phys. Rev. Lett. {\bf 113}, 160403 (2014).
\bibitem{boyd} S. Boyd and L. Vandenberghe, Convex Optimization, Cambridge University Press (2004).
\bibitem{HJW} N. Gisin, Helv. Phys. Acta, \textbf{62}, 363–371 (1989); P. L. Hughston, R. Jozsa, W. K. Wootters, Phys. Lett. A, {\bf 183}, 14 (1993).
\bibitem{CavJonWis2009} E. G. Cavalcanti, S. Jones, H. M. Wiseman and M. Reid, Phys. Rev. A \textbf{80} 032112 (2009).

\bibitem{cvx} M. Grant and S. Boyd. \textit{CVX: Matlab software for disciplined convex programming}, version 2.0 beta. \href{http://cvxr.com/cvx}{\tt{http://cvxr.com/cvx}}, September 2013; \textit{Graph implementations for nonsmooth convex programs}, Recent Advances in Learning and Control (a tribute to M. Vidyasagar), V. Blondel, S. Boyd, and H. Kimura, editors, pages 95-110, Lecture Notes in Control and Information Sciences, Springer, 2008.
\bibitem{qetlab} N. Johnston. \textit{QETLAB: A MATLAB toolbox for quantum entanglement}, version 0.8. \href{http://www.qetlab.com}{\tt{http://www.qetlab.com}}, April 13, 2015. 
\bibitem{attack} A. Ac\'in, D. Cavalcanti, E. Passaro, S. Pironio and P. Skrzypczyk, arXiv:1505.00053 (2015).
\bibitem{Pusey2015} M. F. Pusey, J. Opt. Soc. Am. B \textbf{32}, A56 (2015).

\bibitem{AciMasPir12} A. Ac\'in, , S. Massar and S. Pironio,  Phys. Rev. Lett. \textbf{108}, 100402 (2012).


\end{thebibliography}

\begin{appendix}
\section{Obtaining the SDP for the guessing probability}
In this appendix we will show how to arrive at the SDP (\ref{eq:sdp}) for Eve's guessing probability.

The most general attack that Eve can implement in the case that she is interested in guessing the result of a single measurement ($x = x^*$) of Alice, is to distribute a state $\rho_{\rA\rB\rE}$ to Alice and Bob (keeping a part for herself) on which she will perform a measurement with POVM elements $M_e$, for $e = 0,\ldots, m_\rA-1$, and distribute to Alice a set of measuring devices which implement the POVMs with elements $M_{a|x}$, for $x = 0,\ldots, n_\rA -1$ and $a = 0,\ldots, m_\rA-1$. When Eve obtains outcome $e$ from her measurement she will give this as her guess for the outcome of Alice. Thus, the guessing probability of Eve is given by
\begin{equation}
P_\guess(x^*) = \sum_e \Tr[(M_{a=e|x^*}\otimes M_e) \rho_{\rA\rE}]
\end{equation}
Alice and Bob can however determine the assemblage $\sigma_{a|x}^\obs$ that they hold, (i.e. the set of conditional states prepared for Bob, along with the corresponding probabilities). Thus the optimisation problem we need to solve is given by
\begin{align}\label{e: raw sdp}
\max_{\rho_{\rA\rB\rE}, M_{a|x}, M_e} &\sum_e \Tr[(M_{a=e|x^*}\otimes M_e) \rho_{\rA\rE}] \\
\mathrm{s.t.}\quad &\Tr_{\rA}[(M_{a|x} \otimes \openone_{\rB}) \, \rho_{\rA\rB}] = \sigma_{a|x}^\obs, &\forall a, x\nonumber \\
 &\rho_{\rA\rB\rE}  \succeq 0, \quad \Tr [\rho_{\rA\rB\rE}] = 1 \nonumber \\
 &M_{a|x} \succeq 0,  \forall a,x, \quad \sum_a M_{a|x} = \openone, \forall x \nonumber \\
 &M_{e} \succeq 0,  \forall e \quad \sum_e M_{e} = \openone. \nonumber
\end{align}
Here, the first constraint is the consistency with the observed assemblage, the second constraints demand that $\rho_{\rA\rB\rE}$ is a valid quantum state and the third and fourth constraints that the measurements $M_{a|x}$ and $M_e$ are valid POVMs.

Defining now the joint assemblage for Alice, Bob and Eve,
\begin{equation}
\sigma_{a|x}^e = \Tr_{\rA\rE}[(M_{a|x} \otimes \openone_{\rB} \otimes M_e) \, \rho_{\rA\rB\rE}],
\end{equation}
it is straightforward to see that all of the constraints appearing in (\ref{eq:sdp}) are satisfied whenever the constraints in (\ref{e: raw sdp}) are satisfied, and that the objective functions match. Thus it is straightforward to see that the optimisation problem (\ref{eq:sdp}) is at least a relaxation of (\ref{e: raw sdp}). What we will show now is that they are in fact equivalent optimisation problems by showing that any solution to (\ref{eq:sdp}) also implies a solution to (\ref{e: raw sdp}).

First of all, consider an assemblage $\sigma_{a|x}^e$ satisfying all of the constraints in (\ref{eq:sdp}). For a fixed $e$, we can define $P_\rE(e) = \sum_a \Tr \sigma_{a|x}^e $\footnote{Note that $P_\rE(e)$ is indeed independent of $x$, due to no-signalling, since $\sum_a \sigma_{a|x}^e = \sum_e\sigma_{a|x'}^e$ is independent of $x$.}, and $\tilde{\sigma}_{a|x}^e = \sigma_{a|x}^e/P_\rE(e)$. This has the following properties
\begin{equation}
\begin{split}
&\sum_{a}\tilde{\sigma}_{a|x}^e = \sum_{a}\tilde{\sigma}_{a|x'}^e \quad\forall e, x\neq x', \\ &\Tr\sum_{a}\tilde{\sigma}_{a|x}^e = 1 \,\quad\qquad\forall e
\end{split}
\end{equation}
which show that for each $e$, $\tilde{\sigma}_{a|x}^e$ is a valid assemblage \cite{Pusey}. From the GHJW theorem \cite{HJW} it therefore follows that there is a quantum state $\rho_{\rA\rB}^e$ and POVM elements $M_{a|x}^e$ such that
\begin{equation}
\Tr_\rA[(M_{a|x}^e \otimes \openone_\rB)\rho_{\rA\rB}^e] = \tilde{\sigma}_{a|x}^e
\end{equation}
Now, we finally consider that Eve also sends an additional degree of freedom which is read by the measuring device of Alice -- an auxiliary classical `flag' system, which we label $\rA'$. This system has orthogonal states $\ket{e}$, for $e = 0,\ldots, m_\rA-1$. This system will be read by Alice's measuring device, and, conditioned on the flag, the appropriate measurement will be made. We can thus now  construct the complete strategy of Eve
\begin{align}
\rho_{\rA\rB\rE} &= \sum_e P_\rE(e) \ket{e}\bra{e}_{\rA'} \otimes\rho_{\rA\rB}^e\otimes  \ket{e}\bra{e}_\rE \nonumber \\
M_{a|x} &= \sum_e \ket{e}\bra{e}_{\rA'} \otimes M_{a|x}^e \nonumber \\
M_e &= \ket{e}\bra{e}_\rE
\end{align}
Clearly this defines a valid state and valid measurements, hence they satisfy the latter constraints of (\ref{e: raw sdp}). Furthermore, by construction it also satisfies the first consistency constraint, which is straightforwardly verified.

In total, we thus conclude that the two optimisation problems are equivalent, since the solution to either one implies a solution to the other, obtaining the same $P_\guess(x^*)$. We thus focus on the problem (\ref{eq:sdp}) which is easier to solve, being an SDP optimisation, linear in the optimisation variables $\sigma_{a|x}^e$.

\section{Derivation of the Prepare-and-Measure SDP}\label{a:PM SDP}
In this appendix we will show that the amount of randomness that can be certified in the prepare-and-measure scenario when Alice receives her share of the state through an untrusted channel, and does not trust her measuring device, is given by the SDP (6) in the main text.

Bob prepares a known bipartite state $\rho_{\rA\rB}$ half of which is sent to Alice through the insecure quantum communication channel. Eve can intercept the state, and the most general operation she can perform (in the case that she is guessing only the outcome of a single measurement $x = x^*$) is a measurement with Kraus operators $K_e$, i.e. the POVM elements are $M_e = K_e^\dagger K_e$, and the state prepared by Eve after obtaining outcome $e$ is
\begin{equation}
\rho_{\rA\rB}^e = \frac{(K_e \otimes \openone)\rho_{\rA\rB}(K_e^\dagger\otimes \openone)}{\Tr[K_e \rho_\rA K_e^\dagger]}
\end{equation}
which occurs with probability $P_\rE(e) = \Tr[M_e \rho_\rA]$. Eve will guess that the outcome of Alice's measurement is $e$. Eve now forwards the state onto Alice, and since she controls completely Alice's device, she will allow the device to perform the measurement $N_{a|x}^e$ when her outcome was $e$, and when Alice chooses to make measurement $x$ (that is, Eve sends the classical information of which outcome she obtained along with the quantum state). Thus, the probability for Alice to obtain outcome $a$, given that she made measurement $x$ and Eve obtained outcome $e$ is given by
\begin{equation}
P_\rA(a|x,e) = \frac{\Tr[N_{a|x}^e K_e \rho_\rA K_e^\dagger]}{\Tr[K_e \rho_\rA K_e^\dagger]}.
\end{equation}
Putting everything together, we see therefore that the guessing probability is given by allowing Eve to optimise over all available strategies, and is given by
\begin{align}\label{eq:PM raw}
P_\guess(x^*)=\max &\sum_{e} \Tr[N_{a=e|x^*}^e K_e \rho_\rA K_e^\dagger] \\
\mathrm{s.t.}&\sum_e \Tr_\rA[(K_e^\dagger N_{a|x}^eK_e\otimes \openone)\rho_{\rA\rB}] = \sigma_{a|x}^\obs  \nonumber \\
& \sum_a N_{a|x}^e = \openone \quad\qquad\qquad\qquad\qquad\forall \, e,x \nonumber \\
& \sum_e K_e^\dagger K_e = \openone  \nonumber \\
& N_{a|x}^e \succeq 0 \qquad\qquad\qquad\qquad\qquad\forall \, a,e,x \nonumber
\end{align}
Currently, this optimisation is not in the form of an SDP, due to the nonlinear nature of the objective function and the constraints. However, it can easily be written in the form of an SDP by introducing the new variable $M_{a|x}^e = K_e^\dagger N_{a|x}^e K_e$. The three final constraints on $N_{a|x}^e$ and $K_e$ imply the following constraints on $M_{a|x}^e$,
\begin{align}
\sum_a M_{a|x}^e &= \sum_a M_{a|x'}^e,& &\forall \, e,x'\neq x, \nonumber \\
\sum_{ae} M_{a|x}^e &= \openone,& &\forall \, x, \\
M_{a|x}^e &\succeq 0,&  &\forall \, a,e,x. \nonumber
\end{align}
However, we can see that whenever we have a set of $M_{a|x}^e$ satisfying the above constraints, it implies that there exist $N_{a|x}^e$ and $K_e$ satisfying the original constraints -- i.e. the two sets are equivalent. To see this, we denote first $M_e = \sum_a M_{a|x}^e \succeq 0$ (which is independent of $x$), and therefore we can write $M_e = K_e^\dagger K_e$, for some $K_e$, which is always possible for a positive semi-definite operator. Moreover, since $\sum_{ae} M_{a|x}^e = \sum_e K_e^\dagger K_e = \openone$, the second constraint is satisfied. Finally, defining $N_{a|x}^e = (K_e^\dagger)^{-1}M_{a|x}^e(K_e)^{-1} \succeq 0$ (using the pseudo-inverse when necessary), we also have that
\begin{align}
\sum_a N_{a|x}^e &= (K_e^\dagger)^{-1} M_e (K_e)^{-1} \nonumber \\
&= (K_e^\dagger)^{-1} K_e^\dagger K_e (K_e)^{-1} = \openone
\end{align}
Thus, we can re-express the optimisation problem (\ref{eq:PM raw}) in the form of the following SDP
\begin{align}\label{eq:PM raw2}
P_\guess(x^*)=\max_{M_{a|x}^e}& \sum_{e} \Tr[M_{a=e|x^*}^e \rho_\rA] \\
\mathrm{s.t.}& \sum_e \Tr_\rA[(M_{a|x}^e\otimes \openone)\rho_{\rA\rB}] = \sigma_{a|x}^\obs  \quad\forall \, a,x\nonumber \\
& \sum_a M_{a|x}^e = \sum_a M_{a|x'}^e \qquad\quad\,\,\forall \, e,x\neq x' \nonumber \\
& \sum_{ae} M_{a|x}^e = \openone  \qquad\qquad\qquad\qquad\quad\,\,\,\,\forall \, x \nonumber \\
& M_{a|x}^e \succeq 0 \qquad\qquad\qquad\qquad\qquad\forall \, a,e,x \nonumber
\end{align}
which is exactly the optimisation problem given in the main text. 

\section{Deriving the dual of the SDP (\ref{eq:sdp})}\label{a:dual}
In this appendix we show the explicit form of the dual of the SDP (\ref{eq:sdp}), and explain why Eq. (\ref{eq:dual}) is an equivalent form, which is easier to interpret.

As a reminder, the primal problem is given by
\begin{align}
P_\guess(x^*)=\max_{\sigma_{a|x}^e} & \sum_e \Tr[\sigma_{a=e|x^*}^e]  \\
\mathrm{s.t.}&\sum_e \sigma_{a|x}^e = \sigma_{a|x}^\obs  &\forall \, a,x \nonumber \\
&\sum_a \sigma_{a|x}^e = \sum_a \sigma_{a|x^*}^e &\forall \, e, \, x \neq x^* \nonumber \\
&\sigma_{a|x}^e \succeq 0 &\forall a,x,e.\nonumber
\end{align}
Let us introduce dual variables $F_{a|x}$, $G_{x}^{e}$ and $H_{a|x}^e$, with respect to the first, second and third set of constraints respectively, and form the Lagrangian for this problem,
\begin{multline}
\mathcal{L} = \sum_e \Tr[\sigma_{a=e|x^*}^e] + \sum_{ax} \Tr[F_{a|x}(\sigma_{a|x}^\obs - \sum_e \sigma_{a|x}^e)] \\
+ \sum_{aex} \Tr[G_x^e (\sigma_{a|x}^e - \sigma_{a|x^*}^e)] + \sum_{aex}\Tr[H_{a|x}^e \sigma_{a|x}^e]
\end{multline}
After re-arranging, and grouping terms, this is equivalent to
\begin{multline}
\mathcal{L} = \sum_{ax} \Tr[F_{a|x}\sigma_{a|x}^\obs] 
+ \sum_{aex} \Tr[(\delta_{a,e}\delta_{x,x^*}\openone - F_{a|x} \\+ G_x^e - \delta_{x,x^*}\sum_{x'} G_{x'}^e + H_{a|x}^e )\sigma_{a|x}^e] 
\end{multline}
This Lagrangian provides an upper bound on the primal objective as long as $H_{a|x}^e \succeq 0$. Moreover, it provides a non-trivial upper bound only when the inner bracket in the second line identically vanishes for each value of $a,e,x$. Thus, we arrive at the dual problem
\begin{align}
P_\guess(x^*)=\min& \sum_{ax} \Tr[F_{a|x}\sigma_{a|x}^\obs] \\
\mathrm{s.t.}\,&\delta_{a,e}\delta_{x,x^*}\openone - F_{a|x} + G_x^e \nonumber \\
&- \delta_{x,x^*}\sum_{x'} G_{x'}^e + H_{a|x}^e = 0  &\quad\forall \, a,e,x \nonumber \\
& H_{a|x}^e \succeq 0 &\quad\forall \, a,e,x \nonumber
\end{align}
However, $H_{a|x}^e$ is playing the role of a slack variable, since it doesn't appear in the objective function, so we can finally simplify the dual to arrive at
\begin{align}\label{e:dual sdp}
P_\guess(x^*)=\min_{F_{a|x},G_x^e} &\sum_{ax} \Tr[F_{a|x}\sigma_{a|x}^\obs] \\
\mathrm{s.t.}\quad &F_{a|x} - \delta_{a,e}\delta_{x,x^*}\openone - G_x^e \nonumber \\
&+ \delta_{x,x^*}\sum_{x'} G_{x'}^e \succeq 0  &\quad\forall \, a,e,x \nonumber
\end{align}
The dual is easily seen to be strictly feasible, for example by taking $G_x^e = 0$ and $F_{a|x} = \alpha \openone$ for $\alpha > 1$. Thus strong duality holds, and the optimal value of the dual is equal to the optimal value of the primal. In the form (\ref{e:dual sdp}), the dual is seen manifestly to be an SDP, as expected. Finally, to understand the meaning of the constraint, we multiply by an arbitrary valid assemblage $\sigma_{a|x}$, and take the sum in $a$ and $x$ and the trace. We find
\begin{equation}
\sum_{ax}\Tr[F_{a|x}\sigma_{a|x}] \geq \Tr[\sigma_{e|x^*}] = P(e|x^*)
\end{equation}
must hold for all $e$. Since this condition also holds for all valid assemblages, we see that the second constraint enforces that the value of the inequality is a uniform upper bound on the probability that any individual outcome occurs for the measurement $x^*$, independent of the assemblage. Hence, one sees immediately why this bounds the guessing probability.

\section{Maximal Randomness from all pure states}
In this section we will show analytically that appropriate measurements on all partially entangled qudit states necessarily lead to 1 dit of randomness.

Consider first the partially entangled two qubit state in Schmidt form, $\ket{\psi} = \cos\theta \ket{00} + \sin\theta\ket{11}$, for $\theta \in (0,\pi/4]$, and that Alice's two measurements are $X$ and $Z$ measurements respectively. The assemblage created for Bob is then
\begin{eqnarray}
\sigma_{0|0} &=& \frac{1}{2}\proj{\uparrow_\theta}, \nonumber \\ 
\sigma_{1|0} &=& \frac{1}{2}\proj{\uparrow_{-\theta}}, \nonumber \\  \sigma_{0|1} &=& \cos^2\theta\proj{0}, \nonumber \\ 
\sigma_{1|1} &=& \sin^2\theta\proj{1}, 
\end{eqnarray}
where $\ket{\uparrow_\theta} = \cos\theta\ket{0}+\sin\theta\ket{1}$. Crucially, each element of the assemblage is pure, i.e. each element is of the form $\sigma_{a|x} = P(a|x)\Pi_{a|x}$, where $\Pi_{a|x}$ is a one-dimensional projector. The purity of Bob's assemblage substantially constrains Eve's possible strategies, such that
\begin{equation}
\sigma_{a|x}^e = q(ae|x)\Pi_{a|x}
\end{equation}
where each $q(ae|x) \geq 0$. This says that Eve must prepare the same pure state for Bob in each instance, all she can vary is the probability of the two outcomes (which must still be positive). To be consistent with the observed assemblage, we must have that 
\begin{equation}
\sum_e q(ae|x) = P(a|x).
\end{equation}
The guessing probability also now becomes
\begin{equation}
P_g = \sum_e \tr[\sigma_{a=e|0}^e] = q(00|0) + q(11|0).
\end{equation}

Now, the no-signalling constraint says that $\sum_a \sigma_{a|0}^e = \sum_a \sigma_{a|1}^e$ for all $e$. Specifically, in the case at hand
\begin{equation}
q(0e|0) \Pi_{0|0} + q(1e|0) \Pi_{1|0} = q(0e|1) \Pi_{0|1} + q(1e|1) \Pi_{1|1}, 
\end{equation}
which must be true for all matrix elements. While the projectors on the right-hand-side, corresponding to measurements of $Z$, are diagonal, the left-hand-side, corresponding to $X$, are in general not diagonal. Thus, taking the trace with $\ketbra{0}{1}$, we arrive at the condition
\begin{equation}
\cos\theta\sin\theta(q(0e|0) - q(1e|0)) = 0.
\end{equation} 
Since $\cos\theta\sin\theta\neq 0$ for $\theta \in (0,\pi/4]$, this implies that $q(0e|0) = q(1e|0)$. In particular, this says that $q(01|0) = q(11|0)$. However, to be consistent $q(00|0) + q(01|0) = p(0|0) = 1/2$, and thus we arrive at
\begin{equation}
1/2 = q(00|0) + q(01|0) = q(00|0) + q(11|0) = P_g.
\end{equation}
Thus, analytically it must be the case that $P_g = 1/2$, and hence 1 bit of randomness is obtained by measuring $X$ and $Z$ on any partially entangled state of two qubits.

The above also extends to qudits; assuming that the state has Schmidt-rank $d$ then 1 dit of randomness can always be obtained. Let us now write the state as
\begin{equation}
\ket{\psi} = \sum_{k=0}^{d-1}\sqrt{\lambda_k}\ket{k}\ket{k}
\end{equation}
where $\sum_k \lambda_k = 1$, and $\lambda_k > 0$. Alice's first measurement will now be in the Fourier transform basis, with eigenstates
\begin{equation}
\ket{\tilde{a}} = \frac{1}{\sqrt{d}} \sum_{k=0}^{d-1}\omega^{ak}\ket{k}
\end{equation}
and $\omega = e^{2\pi i/d}$ the corresponding root of unity. Her second measurement will be in the $Z$ basis with eigenstates $\{\ket{a}\}$. For Alice's first measurement she obtains each outcome with equal probability $P(a|0) = 1/d$, and prepares the pure states for Bob $\Pi_{a|0}$, given by
\begin{equation}
\Pi_{a|0} = \sum_{kl}\sqrt{\lambda_k \lambda_l} \omega^{a(l-k)}\ket{k}\bra{l}.
\end{equation}
For Alice's second measurement, she obtains outcome $a$ with probability $P(a|1) = \lambda_a$, and prepares the state $\Pi_{a|1} = \ket{a}\bra{a}$. As above, the purity of Bob's assemblage means that Eve is again forced to use strategies of the form $\sigma_{a|x}^e = q(ae|x)\Pi_{a|x}$. For consistency we still have $\sum_e q(ae|x) = P(a|x)$, for the guessing probability $P_g = \sum_e q(ee|0)$, and from no-signalling $\sum_a q(ae|0)\Pi_{a|0} = \sum_a q(ae|1)\Pi_{a|1}$. Once again, the right-hand-side is diagonal, and hence by looking at the off-diagonal matrix elements, i.e. by taking the trace with $\ket{k}\bra{l}$, we find that
\begin{equation}
\sum_a q(ae|0)\sqrt{\lambda_k\lambda_l} \omega^{a(l-k)} = 0
\end{equation}
Since, by assumption of being Schmidt-rank $d$, none of the Schmidt coefficients vanish, we therefore must have that 
\begin{equation}
\sum_a q(ae|0)\omega^{a(l-k)} = 0.
\end{equation}
Considering only the elements with $k = 0$ (and $l = 1,\ldots, d-1$), along with the equation $\sum_a q(ae|0) = P(e)$, which says that Eve's probability to output $e$ is just the marginal distribution, we notice that this set of equations, when combined, has the familiar form of a discrete Fourier transform (up to normalisation): 
\begin{equation}\left[
\begin{array}{cccc}
    1  &  1     & \dots & 1 \\
    1  & \omega & \dots & \omega^{d-1} \\
    \vdots & \vdots & \ddots & \vdots \\
    1  & \omega^{d-1} & \dots & \omega^{(d-1)^2}
\end{array}\right]
\left[
\begin{array}{c}
q(0e|0) \\ q(1e|0) \\ \vdots \\ q(d-1,e|0)
\end{array}\right] = 
\left[
\begin{array}{c}
P(e) \\ 0 \\ \vdots \\ 0
\end{array}\right]
\end{equation}
Thus, this equation is readily inverted, and we obtain as solution $q(ae|0) = P(e)/d$ for all $a, e$. In particular, this implies that Eve's guess is completely uncorrelated from Alice's, and her guessing probability is $P_g = \sum_e q(ee|0) = \frac{1}{d}\sum_e P(e) = 1/d$. Thus 1 dit of randomness is obtained from Alice's measurement. 

\end{appendix}
\end{document}